\documentclass[10pt, conference]{IEEEtran}
\usepackage{algorithmicx}
\usepackage[ruled,vlined,linesnumbered]{algorithm2e}
\usepackage{color}
\usepackage{hhline}
\usepackage{amsmath,mathtools}
\usepackage{amsfonts,amssymb}
\usepackage{mathrsfs}
\usepackage{gensymb} 
\usepackage{caption}
\usepackage{multirow}
\usepackage{graphicx} 
\usepackage{multirow}
\usepackage{enumitem,color}
\usepackage{algpseudocode}

\begin{document}
%
\title{\huge \emph{oneTwin}: Online Digital Network Twin via Neural Radio Radiance Field\vspace{-0.15in}}

\author{\IEEEauthorblockN{Yuru Zhang, Ming Zhao, Qiang Liu \vspace{-0.16in}}\\
\IEEEauthorblockA{
University of Nebraska-Lincoln\\
yzhang176@huskers.unl.edu}\vspace{-0.4in}
\and
\IEEEauthorblockN{Nakjung Choi \vspace{-0.16in}}\\
\IEEEauthorblockA{
Nokia Bell Labs\\
nakjung.choi@nokia-bell-labs.com}\vspace{-0.4in}
}

\maketitle

\begin{abstract}
Digital network twin is a promising technology that replicates real-world networks in real-time and assists with the design, operation, and management of next-generation networks.
However, existing approaches (e.g., simulator-based and neural-based) cannot effectively realize the digital network twin, in terms of fidelity, synchronicity, and tractability.
In this paper, we propose oneTwin, the first online digital twin system, for the prediction of physical layer metrics.
We architect the oneTwin system with two primary components: an enhanced simulator and a neural radio radiance field (NRRF).
On the one hand, we achieve the enhanced simulator by designing a material tuning algorithm that incrementally optimizes the building materials to minimize the twin-to-real gap.
On the other hand, we achieve the NRRF by designing a neural learning algorithm that continually updates its DNNs based on both online and simulated data from the enhanced simulator.
We implement oneTwin system using Sionna RT as the simulator and developing new DNNs as the NRRF, under a public cellular network.
Extensive experimental results show that, compared to state-of-the-art solutions, oneTwin achieves real-time updating (0.98s), with 36.39\% and 57.50\% reductions of twin-to-real gap under in-distribution and out-of-distribution test datasets, respectively. 
\end{abstract}

\begin{IEEEkeywords}
Digital Network Twin, Neural Radio Radiance Field, Online Learning, Simulator Enhancement
\end{IEEEkeywords}

\section{Introduction}
\label{sec:introduction}

Digital network twin (DNT)~\cite{wu2021digital, khan2022digital, nguyen2021digital} is a promising paradigm to replicate live mobile networks in the real world, with essential attributes of fidelity, synchronicity, and tractability.
The fidelity attribute ensures that a DNT can accurately represent the real-world network, including network contexts, state transitions, and performance metrics~\cite{almasan2022network}. 
The synchronicity attribute ensures that DNT can track and reflect the time-evolving dynamics of the real-world network in real time (e.g., subseconds)~\cite{mashaly2021connecting}. 
The tractability attribute guarantees that DNT can be massively queried and inferred at runtime, within a given computational complexity.
DNT is highly anticipated to facilitate and assist a wide range of network operations, management, and maintenance in next-generation networks~\cite{zhang2024digital,sun2020reducing, dai2020deep}. 
For example, consider an autonomous food delivery UAV performing online trajectory planning across buildings in a dense urban scenario. 
A DNT of physical layer metrics (e.g., reference signal received power (RSRP) and MIMO channel) can be tremendously beneficial in ensuring consistently stable wireless connectivity (e.g., throughput and link reliability).

\begin{figure}[!t]
	\centering
	\includegraphics[width=3.45in]{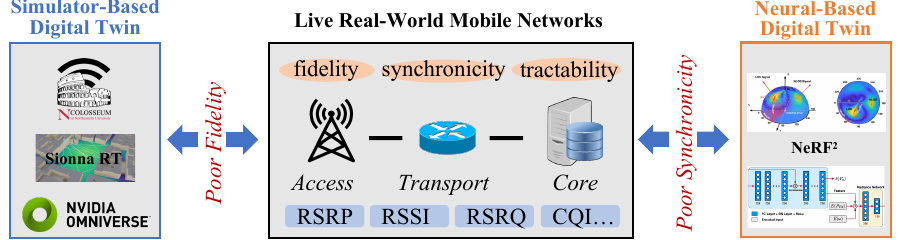}
	  \caption{Existing approaches of digital network twin.}
	\label{fig:digital_twin}
\end{figure}

To achieve the DNT, existing works generally divide into two approaches, as illustrated in Fig.~\ref{fig:digital_twin}.
On the one hand, simulator-based approaches aim to build DNTs using offline network simulators, such as Colosseum~\cite{villa2024colosseum, polese2024colosseum}, BostonTwin~\cite{testolina2024boston}, and CAVIAR~\cite{borges2024caviar}.
The main idea is to replicate the real-world network with identical setups in the simulator, where the simulation parameters are calibrated as precisely as possible.
However, these DNTs can hardly achieve the attribute of synchronicity, as any time-evolving network dynamics need to be updated manually and their inferences in large-scale scenarios are mostly non-real-time due to complicated simulation procedures.
Moreover, recent works~\cite{zhang2020onrl,liu2022atlas, shi2021adapting} have revealed that the simulation-to-reality (sim-to-real) gap is non-trivial (sometimes significant), which compromises the fidelity of these simulator-based DNTs. 
On the other hand, neural-based approaches aim to build DNTs by leveraging the high approximation capability of deep neural networks (DNNs), such as NeRF$^2$~\cite{zhao2023nerf2}.
Their ideas are to collect massive data from real-world networks and train dedicated DNNs to imitate network behaviors~\cite{jia2024neural, jiang2025learnable, zhao2023nerf2}.
By translating the neural radiance field (NeRF) technique~\cite{mildenhall2021nerf} into radio-frequency (RF) signal propagation domain, NeRF$^2$~\cite{zhao2023nerf2} achieves high fidelity (e.g., low prediction errors of received signal strength indicator (RSSI)) and real-time inference (e.g., within subseconds under GPU acceleration).
However, neural-based approaches suffer from data efficiency and generalizability concerns, which limit their capability to online track time-evolving dynamics in real time, especially under unseen states and geographic areas.

In this paper, we propose \underline{o}nline digital \underline{ne}twork \underline{twin} (oneTwin), the \emph{first} online digital network twin system.
The fundamental idea is to learn a neural radio radiance field (NRRF) on-the-fly by augmenting online measurements with simulated data from an enhanced simulator.
The rationale is that simulators embed rigorous domain knowledge, and simulated data can regularize the learning process, improving the generalizability of neural models in real-world networks.
Additionally, we enhance the simulator itself by reducing its sim-to-real gap with respect to the real-world environment.
We architect the oneTwin system with two primary components: an enhanced simulator and the NRRF.
The enhanced simulator is achieved by asynchronously optimizing the radio material properties of buildings in the scene. 
To achieve this, we design a new automatic material tuning algorithm based on Bayesian optimization, that minimizes the sim-to-real gap under accumulated online measurements.
The NRRF is achieved by continually learning and updating its DNNs to adapt to time-evolving dynamics in the real-world network.
To this end, we develop a new online neural learning algorithm, which reduces the twin-to-real gap while ensuring the generalizability of the NRRF, based on both online data and offline simulated data from the enhanced simulator.
Note that, the simulator and NRRF operate in a decoupled manner: the NRRF is updated in real time with sub-second updating latency, while the simulator is enhanced asynchronously to improve the fidelity of simulated data.
We implement the oneTwin system by developing the simulator based on Sionna RayTracing and the NRRF, under a public cellular network.
Experimental results show that compared to state-of-the-art solutions, oneTwin achieves real-time updating (0.98s), with 36.39\% and 57.50\% reductions of twin-to-real gap under in-distribution and out-of-distribution test datasets, respectively.

Overall, we propose oneTwin, the \emph{first} \underline{online} digital network twin system, with the following contributions:
\begin{itemize} [leftmargin=*]
    \item We design an automatic material tuning algorithm that achieves enhanced simulator with reduced sim-to-real gap.
    \item We introduce an online neural learning algorithm that achieves the NRRF with real-time updating, high tractability and improved generalizability to unseen environments.
    \item We implement oneTwin based on Sionna RT simulator with online measurements collected from a public network.
    \item We evaluate oneTwin via extensive experiments, in terms of fidelity, synchronicity, tractability, and generalizability.
\end{itemize}

\section{Problem Statement and System Overview}
\label{sec:overview}

\textbf{Problem Statement.}
The propagation of RF signals in real-world wireless networks is extremely complex, involving various electromagnetic effects such as absorption, reflection, diffraction, and scattering~\cite{yun2015ray}.
For example, reflection occurs when RF signals propagate on certain surfaces (e.g., walls and roofs), depending on material characteristics (e.g., concrete, metal).
These effects interplay in complex and dynamic radio environments (e.g., buildings and vehicles)~\cite{zhao2023nerf2}, making it computationally intractable to derive accurate closed-form models of real-world signal propagation.

In this paper, we focus on the reference signal received power (RSRP)\footnote{The oneTwin system can be extended to support other physical layer metrics simultaneously, such as reference Signal Received Quality (RSRQ), received signal strength indicator (RSSI), and channel quality indicator (CQI).}, a representative physical-layer metric that captures the average received power from a reference signal.
Here, we denote \emph{digital twinning} as the process of creating, updating, and maintaining the digital network twin, based on online data points collected from real-world networks.
The objectives of the digital network twin are three-fold: First, we aim to accurately predict the RSRP of mobile users at arbitrary geographic locations to achieve the attribute of \emph{fidelity}.
Here, the input space includes the 3D location (and orientation if needed), and the output space is the predicted RSRP value.
Second, we aim to update the digital network twin online to track time-evolving dynamics in the real-world network, to achieve the attribute of \emph{synchronicity}. 
Here, online data points (denoted as location-to-RSRP pairs) in the real-world network arrive in a streaming manner, such as periodical UAV reporting.
Third, we aim to achieve real-time querying of the digital network twin (e.g., subseconds) for potentially massive users, with the attribute of \emph{tractability}.
To quantify the accuracy of digital twinning, we define the \emph{twin-to-reality gap}, representing the average absolute error between predicted and measured RSRP values across a set of locations:
\begin{equation}
    G = \frac{1}{|\mathcal{L}|} \sum\limits_{l_n \in \mathcal{L}} {|R_t(l_n) - R_r(l_n)|},
\end{equation}
where $\mathcal{L} = \{l_1, l_2, ...\}$ is the set of locations, and $R_r(l_n)$ and $R_t(l_n)$ are the RSRP value at the location $l_n$ in the digital network twin and the real-world network, respectively.

\begin{figure}[!t]
	\centering
	\includegraphics[width=3.45in]{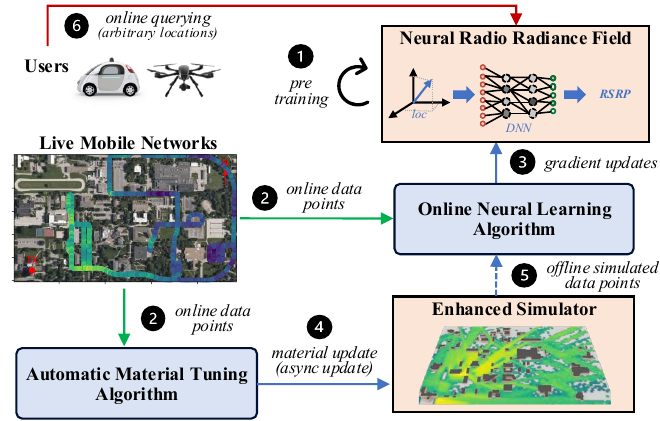}
        \caption{The overview of the oneTwin system.}
	\label{fig:overview}
\end{figure}

\textbf{System Overview.}
Fig.~\ref{fig:overview} overviews the oneTwin system, which consists of two primary components: an enhanced simulator and a neural radio radiance field (NRRF).
The goal is to realize an online digital twin of a live mobile network that enables real-time prediction of physical-layer metrics (i.e., RSRP) at sub-second latency.
The oneTwin system works in the following steps.
First, a network simulator is created to replicate the network deployment (e.g., base station and buildings) of the live mobile network, with calibrated simulation parameters.
We generate extensive simulated data points from the calibrated simulator to pretrain the NRRF, which serves as the initialization of the digital network twin before online adaptation begins.
Second, oneTwin continuously receives online measurements from users in the live network, such as UAVs or vehicles.
Third, the NRRF is continually updated via the neural learning algorithm that adapts its parameters based on both the real-world measurements and the simulated data from the enhanced simulator (see Sec.~\ref{sec:twinning}).
Fourth, the enhanced simulator is asynchronously updated via the material tuning algorithm that optimizes the building material parameters to minimize the sim-to-real gap under all to-date online data points (see Sec.~\ref{sec:augmentation}).
With the asynchronous simulator enhancement, we can accelerate the oneTwin system to be real-time (subseconds), by hiding the non-trivial latency of the material tuning algorithm\footnote{The asynchronous simulator enhancement leads to slightly delayed simulated data (in terms of tracking real-world dynamics), though the resulting performance degradation in the twin-to-real gap remains minimal (see Fig.~\ref{fig:core_sample_efficiency}).}. 
%
Fifth, we selectively generate simulated data from the enhanced simulator, which will be incorporated into the NRRF updates.
Steps 3 through 5 repeat as new data arrives, allowing oneTwin to continuously track and adapt to the time-varying real-world network.
Finally, users can query the NRRF at any time to infer RSRP values at arbitrary locations.

\section{Design of Enhanced Simulator}
\label{sec:augmentation}


\subsection{Problem Formulation}
\label{subsec:augmentation-problem}

Given a network simulator (e.g., Sionna RT), we replicate the target scenario in the real-world network, including network topology (e.g., location and orientation of antennas), configuration (e.g., operating frequency and mode), and parameters (e.g., antenna gain, transmit power, and losses).
We denote the input space as the 3D location (and potential orientation) of the receiver, and the output space as the generated RSRP value.
To balance fidelity and tractability, simulators generally apply various mechanisms of abstraction and simplification~\footnote{Given simulators with various abstraction and approximation, the sim-to-real gap is observed to be non-trivial in practice and difficult to eliminate~\cite{shi2021adapting,liu2021onslicing}. In this work, we mainly focus on parameter tuning to reduce the gap, which is compatible with other applicable approaches.}.
For example, Sionna RT~\cite{hoydis2023sionna} uses the Fibonacci method by default, rather than exhaustive searching, to reduce simulation time in ray-tracing computation.
Since objects in the scenario cannot be exactly matched with those in the real world (e.g., trees and vehicles), the sim-to-real gap exists and is reflected in the difference between the simulated and measured RSRPs.

\begin{figure}[!t] 
\captionsetup{justification=centering}
  \begin{minipage}[t]{0.24\textwidth}
    \centering
    \includegraphics[width=1.7in]{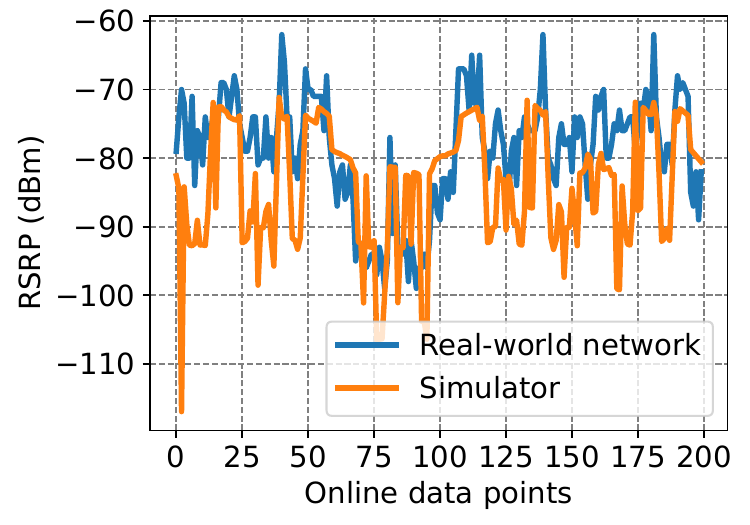}
    \caption{RSRP in simulator and real-world network.}
    \label{fig:sys_sim_rsrp_compare}
  \end{minipage}
  \begin{minipage}[t]{0.24\textwidth}
    \centering
    \includegraphics[width=1.7in]{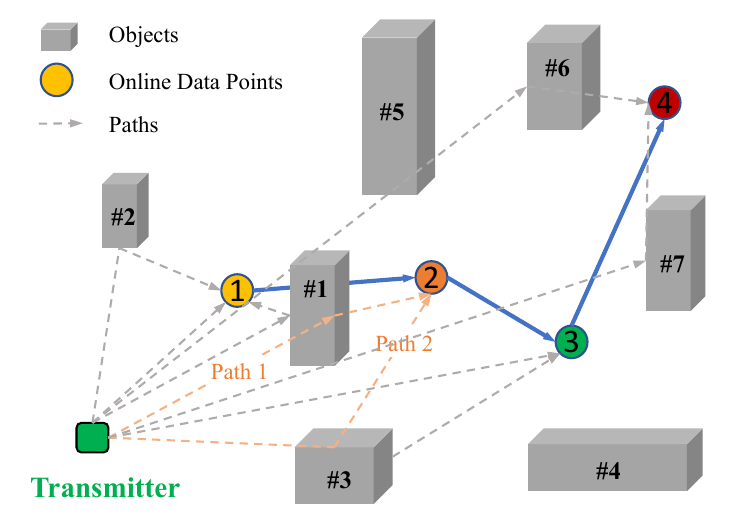}
    \caption{An illustration of RF signal reflection in a scenario.}
    \label{fig:simulator_example_figure}
  \end{minipage}
\end{figure}

To reduce the sim-to-real gap, common practices focus on the calibration of the scenario (i.e., topology, configuration, and parameters), such as antenna profiling and loss measurement~\cite{stetsko2011calibrating}. 
In the oneTwin system, we extensively calibrate the scenario in the simulator via parameter measurement and open database searching (see Sec.~\ref{sec:evaluation}).
Fig.~\ref{fig:sys_sim_rsrp_compare} shows the predicted RSRPs in the calibrated simulator and actual measured RSRPs in the real-world network, during a driving test in a university campus.  
We found that the remaining sim-to-real gap is non-trivial (7.81dB on average), where the maximum RSRP difference can be up to 46.96dB.


Therefore, we propose to further enhance the calibrated simulator to reduce its sim-to-real gap (i.e., the RSRP difference between the simulator and the real-world network).
The fundamental idea is to tune the material properties of objects (especially buildings) in the calibrated simulator, based on accumulated online measurements from the real-world network.
The rationale is that, the material type of objects is critical (e.g., permittivity and conductivity) in computing these electromagnetic effects in the simulator.
Given the massive objects in the scenario (hundreds if not more), it is impractical to accurately identify the exact material composition of all surfaces (e.g., window glass, frame metal).
This uncertainty introduces discrepancies between the simulator and the real-world network.

Denote $m_i$ as the material of the $i$-th object in the scenario and $\mathcal{M} = \{m_i, \forall i \in \mathcal{I}\}$ as the material set, where $\mathcal{I}$ is the set of all objects.
We define $\mathcal{S} = \{itu\_1, itu\_2, ...\}$ as the material space, where $m_i$ can select any material from this space.
Given the material set $\mathcal{M}$, the simulator can be executed to generate the RSRP value at any location in the scenario.
Denote $R_s(l_n|\mathcal{M})$ as the simulator RSRP at the location $l_n$, depending on the selected material set for all objects.
Therefore, given the set of to-date online data points ($\mathcal{L}$), we formulate the material tuning problem as
\begin{align}
  \mathbb{P}_0: \;\;\;\;  \min\limits_{\mathcal{M}} \;\;\;\; & \frac{1}{|\mathcal{L}|} \sum\limits_{l_n \in \mathcal{L}} {|R_s(l_n|\mathcal{M}) - R_r(l_n)|},\\
    s.t. \;\;\;\; & m_i \in \mathcal{S}, \;\; \forall m_i,
\end{align}
where the objective is to minimize the sim-to-real gap by finding the optimal material set.
In the oneTwin system, we keep receiving online data points (i.e., location-to-RSRP pairs), which arrive in the manner of streaming (e.g., one-by-one and batch-by-batch).
In other words, the optimization space (i.e., the dimension of material space) of problem $\mathbb{P}_0$ increases along with the total number of online data points.


The challenges of addressing the above material tuning problem are two-fold.
First, the predicted RSRP of simulator $R_s(l_n|\mathcal{M})$ relates to a wide range of factors and involves complex ray tracing during the simulation.
As a result, it is impossible to mathematically derive its closed-form representation, which results in unknown functions in $\mathbb{P}_0$ and thus falls into the realm of blackbox optimization~\cite{larson2019derivative,conn2009introduction}.
Second, the material space $\mathcal{S}$ is high-dimensional (e.g., tens of candidate materials) and also categorical, where material types have no explicit relations with each other.
Considering the massive number of objects in the scenario, the combinatorial optimization space is tremendous and also ever-increasing.

\subsection{The Solution}
\label{subsec:augmentation-solution}

\begin{figure}[!t]
	\centering
	\includegraphics[width=3.45in]{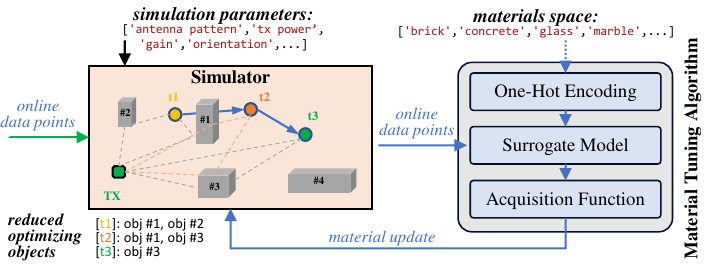}
	\caption{The proposed material tuning algorithm.}
	\label{fig:material_tuning}
\end{figure}

In this subsection, we propose a new automatic material tuning algorithm in Fig.~\ref{fig:material_tuning} to efficiently solve the above problem.
Note that, we are handling the online problem, where new data points come one-by-one (or batch-by-batch) from real-world measurements.
First, we dynamically reduce the problem by freezing the material of certain objects, which significantly decreases the combinatorial optimization space and accelerates the problem-solving.   
Second, we use the Bayesian optimization framework to efficiently handle the blackbox function $R_s(l_n|\mathcal{M})$ and iteratively search for the optimal material set. 
Third, we apply the one-hot encoding technique to convert categorical material space into a binary one, which facilitates the approximation of the surrogate model in Bayesian optimization.

\textbf{Problem Reduction.}
In the problem $\mathbb{P}_0$, the function of simulator RSRP $R_s(l_n|\mathcal{M})$ relates to the material of all objects, which leads to extremely high-dim optimization space under large-scale scenarios (e.g., hundreds of buildings). 
Note that, the simulator can generate all ray paths and involved objects after the ray tracing computation, for arbitrary receivers and locations.
Hence, we can leverage the result of ray tracing (i.e., the involved objects at the current location) to reduce the combinatorial optimization space, by only focusing on the involved objects, rather than all objects.
Although the ray tracing computation may be simplified (possibly inaccurate), these results generally represent the domain knowledge regarding the electromagnetic effects inside the scenario in the real-world network.
For example, Fig.~\ref{fig:simulator_example_figure} illustrates the reflection of RF signals among different objects.
For the first online data point, there is one line-of-sight path and two reflection paths, involving object 1 and object 2.
The material tuning problem could only focus on object 1 and object 2, and keep the material of other objects fixed, because their materials have negligible impact on the simulator RSRP at the first data point.
Likewise, we could focus on tuning the material for object 1 and object 3, when the second online data point arrives.

Hence, we reduce the problem $\mathbb{P}_0$ by only optimizing the involved objects for the latest online data point, while keeping the material of other objects fixed, i.e.,
\begin{align}
  \mathbb{P}_1: \;\;\;\;  \min\limits_{\mathcal{K}} \;\;\;\; & \frac{1}{|\mathcal{L}|} \sum\limits_{l_n \in \mathcal{L}} {|R_s(l_n|\mathcal{K}) - R_r(l_n)|},\\
    s.t. \;\;\;\; & m_i \in \mathcal{S}, \;\; \forall m_i \in \mathcal{K},
\end{align}
where $\mathcal{K} \subseteq \mathcal{M}$ is the set of objects that only relates to the latest online data point.
To further reduce the optimization space, we rank all the involved objects $\mathcal{K}$ according to their strengths, and remove insignificant paths and corresponding objects.
Note that, these objects $\mathcal{K}$ could be involved in the ray tracing computation of previous online data points, e.g., the ray paths of both data point 1 and 2 include object 1, in Fig.~\ref{fig:simulator_example_figure}.
In other words, tuning the material of objects $\mathcal{K}$ may also affect the simulator RSRPs of previous online data points, and thus the function of their simulator RSRPs cannot be further reduced in the above problem $\mathbb{P}_1$.

\textbf{Categorical Bayesian Optimization.}
Next, we develop a categorical Bayesian optimization to resolve the reduced problem $\mathbb{P}_1$ with blackbox functions.
Bayesian optimization~\cite{frazier2018tutorial, wang2023recent, shahriari2015taking} is a state-of-the-art global optimization framework, known for its high sample efficiency and robustness in solving various blackbox problems in practice. 
It relies on a surrogate model to approximate the blackbox function and an acquisition function to guide the selection of subsequent evaluation actions. 
In each iteration, the surrogate model is updated with new observations to improve its prediction accuracy, and then the acquisition function identifies the next action with the best utility, by balancing the exploration and exploitation.

\emph{\underline{Surrogate Model}.} We employ Gaussian Processes (GP) as the surrogate model to approximate the blackbox function $f(\mathcal{K})=\sum\nolimits_{l_n \in \mathcal{L}} {|R_s(l_n|\mathcal{K}) - R_r(l_n)|}$. 
GP extends the concept of the multivariate Gaussian distribution to infinite dimensions, expressed as $ f(\mathcal{K}) \sim \mathcal{G}\mathcal{P}\left(\mu(\mathcal{K}), {c}\left(\mathcal{K}, \mathcal{K}^{\prime}\right)\right)$. 
This formulation represents the distribution of functions $f(\mathcal{K})$ through a mean function $\mu(\mathcal{K})$ and a covariance function ${c}\left(\mathcal{K}, \mathcal{K}^{\prime}\right) $. The covariance within a GP is defined by a kernel function, and in this case, we use the widely adopted Radial Basis Function (RBF) kernel, specified as ${c}\left(\mathcal{K}, \mathcal{K}^{\prime}\right) = \sigma^2 \exp \left(-{\left\|\mathcal{K} - \mathcal{K}^{\prime}\right\|^2}/{2 d^2}\right)$. Here, the hyperparameters $\sigma$ (variance) and $d$ (distance) control the average deviation from the mean function and the range of influence on nearby actions, respectively, and $\left\|\mathcal{K} - \mathcal{K}^{\prime}\right\|^2$ represents the distance between two actions.

\emph{\underline{Acquisition Function}.} 
We adopt expected improvement (EI) as the acquisition function, which shows high robustness in a variety of real-world application domains~\cite{wilson2018maximizing,de2019sampling}. 
With Gaussian process as the surrogate model, EI is expressed as
\begin{equation}
\text{EI}(\mathcal{K})= \left(\mu(\mathcal{K})-f\left(\mathcal{K}^{+}\right)-\epsilon\right) \Phi(Z)+\sigma(\mathcal{K}) \phi(Z),
\end{equation}
if $\sigma(\mathcal{K})>0$, otherwise $\text{EI}(\mathcal{K}) = 0$.
Here, $f\left(\mathcal{K}^{+}\right)$ is the best value in previous observations, and 
\begin{equation}
Z={(\mu(\mathcal{K})-f\left(\mathcal{K}^{+}\right)-\epsilon)}/{\sigma(\mathcal{K})},
\end{equation}
where $\mu(\mathcal{K})$ and $\sigma(\mathcal{K})$ are the mean and the standard deviation of the GP at $\mathcal{K}$, respectively. $\Phi$ and $\phi$ are the CDF and PDF of the standard normal distribution, respectively. 
Here, $\epsilon$ is a small non-negative constant to balance the exploration and exploitation.
By maximizing the above EI acquisition function over the feasible optimization space, the next material set $\mathcal{K}$ will be selected in Bayesian optimization.

\emph{\underline{One-Hot Encoding}.} 
Then, we convert the categorical material space into a binary one by using one-hot encoding~\cite{rodriguez2018beyond}.
The basic idea is to build a binary vector, in which each position corresponds to one category in the material space.
For example, consider there is only one object in the scenario and its material type can be selected from all $B$ possible material types $\{ m_1, m_2, ..., m_B \}$. 
We define a one-hot encoding function to convert the categorical material type $m_i$ into a binary vector of length $B$. 
Specifically, for the material type $m_b$ (where $b$ is between $1$ and $B$), the one-hot encoded vector has the value 1 at the $b$-th position and 0 at all other positions. 
With the converted binary material space, the surrogate model can be efficiently applied to approximate the blackbox function.





\section{Design of Neural Radio Radiance Field}
\label{sec:twinning}

\subsection{Problem Formulation}
\label{subsec:twinning-problem}
With the aforementioned material tuning algorithm, we achieve the enhanced simulator with a reduced twin-to-real gap.
However, the enhanced simulator can hardly be the digital network twin, because 1) its high computation complexity leads to non-real-time online querying, especially for large-scale scenarios; 2) the remaining sim-to-real gap cannot be further minimized with online data points, limited by its abstraction and approximation mechanisms.

In the oneTwin system, we design a neural radio radiance field (NRRF) to be the digital network twin, which will be updated online and can be queried in real-time under GPU acceleration.
The NRRF is inspired by NeRF$^2$~\cite{zhao2023nerf2} that translates the NeRF technique from optics into RF domain.

To model the radiance field, the scene is discretized into 3D voxels, and each voxel retransmits the RF signal it receives, becoming a new radiance source. 
We use $P_x$ to denote the position of voxel $x$. The voxel receives RF signals from multiple paths and retransmits the signal along a new path to the RX. 
Each voxel is characterized by its position $P_x = (X, Y, Z)$, attenuation $\delta(P_x)$, and retransmitted RF signal $S(P_x, \omega)$. 
At the same time, voxel has the uneven radiation direction angle, defined by the azimuthal and elevation angles $\omega = (\alpha, \beta)$. NRRF predicts the RF signal $S$ retransmitted from voxel $P_x$ toward direction $\omega$ given the TX's position $P_{TX}$. Hence, the radiance field $\mathcal{F}$~\cite{zhao2023nerf2} is modeled as:
\begin{equation}
\mathcal{F}: (P_{TX}, P_x, \omega) \to (\delta(P_x), S(P_x, \omega)). 
\label{eq:nrrf}
\end{equation}
This field outputs the voxel's attenuation $\delta(P_x)$ and the retransmitted RF signal $S(P_x, \omega)$.
Similar to NeRF$^2$~\cite{zhao2023nerf2}, we predict the distribution of RF signals in the scene, with two deep neural networks (DNNs): \emph{attenuation network} and \emph{radiance network}. 
The attenuation network predicts the attenuation $\delta(P_x)$, which then is combined with the RX direction $\omega$ and TX position $P_{TX}$, and passed to the radiance network to generate retransmitted RF signal $S(P_x, \omega)$. 

In addition to considering the distribution of RF signals, we also need to track it in all potential directions to know what signal is received at the RX. 
The received signal $R$ at the RX is a combination of signals transmitted from all possible directions.
The points on the ray which starts from RX and directs toward $\omega$ can be expressed as $P(r,\omega)=P_{RX}+r\omega, P_{RX} = P(0,\omega)$, where $r$ is the radial distance from the RX to the point on the ray.
Hence, the received signal at the RX from direction $\omega$ is the accumulation of RF signals emitted from all voxels along the ray:
\begin{equation}
R(\omega)=\int_0^{D_{max}} \delta_{P(r, \omega) \rightarrow P_{\mathrm{RX}}} S(P(r, \omega),-\omega) \mathrm{d} r,
\end{equation}
where $S(P(r, \omega),-\omega)$ is the signal transmitted from the voxel $P(r,\omega)$ on the ray to $P_{RX}$, which has the opposite direction to the ray. Here, $D_{max}$ is the maximum distance in the scenario.

The challenges of achieving online NRRF are two-fold.
First, online data points arrive in the manner of streaming (e.g., point-by-point), which falls into the realm of continual learning.
Updating the NRRF with streaming online data points only may lead to catastrophic forgetting, which reduces its generalizability under unseen geographic areas.
Second, at the very beginning of online learning, the extremely limited data points (tens if not less) could result in a very unstable and unpredictable twin-to-real gap of the NRRF.

\subsection{The Solution}
\label{subsec:twinning-solution}

\begin{figure}[!t]
	\centering
	\includegraphics[width=3.45in]{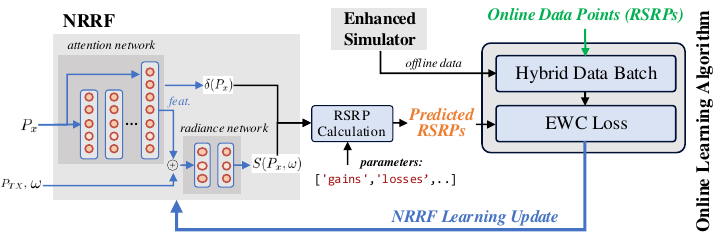}
	\caption{The proposed NRRF and online learning algorithm.}
	\label{fig:NRRF}
\end{figure}

In this subsection, we propose an online neural learning algorithm in Fig.~\ref{fig:NRRF} to efficiently address the aforementioned challenges.
First, we develop a new loss function by introducing an additional penalization term with the elastic weight consolidation (EWC) technique, which penalizes the NRRF training process if the previously learned knowledge is modified.
Second, we build a hybrid data batch, including not only online data points but also simulated data points from the enhanced simulator, which augments limited online data and thus effectively improves its generalizability.
Third, we extensively sample offline data points from the original simulator and use it to train the NRRF offline thoroughly, which serves as the starting point for online digital twinning.

\textbf{Elastic Weight Consolidation.} Elastic weight consolidation (EWC)~\cite{aich2021elastic, kirkpatrick2017overcoming} is an effective technique to mitigate catastrophic forgetting during the continual learning of DNNs. 
In oneTwin, the catastrophic forgetting of NRRF mostly appears under other geographic areas than the proximal areas of current online data points (e.g., flying trajectory of UAVs).
Denote $\theta$ as the parameters of the NRRF, we calculate the Fisher information matrix to measure the information that a given dataset carries about the network parameter $\theta$.
Specifically, it is defined as the covariance of the gradient of the log likelihood estimate of the given dataset $\mathcal{D}$, i.e., 
\begin{align}
F(\theta)=\underset{\substack{\mathbf{x} \sim \mathcal{D} \\ \hat{y} \sim p_{\boldsymbol{\theta}}(y \mid \mathbf{x})}}{\mathbb{E}}\left[\nabla_{\boldsymbol{\theta}} \log p_{\boldsymbol{\theta}}(\hat{y} \mid \mathbf{x}) \nabla_{\boldsymbol{\theta}} \log p_{\boldsymbol{\theta}}(\hat{y} \mid \mathbf{x})^T\right]
\end{align}
where $\mathbf{x}$ and $\hat{y}$ are the inputs of the given dataset and outputs of the NRRF, respectively.
To prevent significant changes to weights that are crucial for previously learned knowledge (e.g., other geographic areas), EWC introduces a regularization term into the loss function. 
This term penalizes the deviation of certain weights according to the above Fisher information matrix. 
Specifically, we define the overall loss function of NRRF during continual learning as
\begin{equation}
Loss = L_{MSE} + \frac{\lambda}{2} \sum\nolimits_i F_{i} (\theta_i - \theta_i^*)^2, 
\end{equation}
where $L_{MSE}$ is the original MSE training loss function of NRRF (see Eq.~\ref{eq:nrrf}), $\theta_i^*$ is the last learned $i$th weight, and $\lambda$ is a hyperparameter to control the enforcement of the penalization. 

\textbf{Hybrid Data Batch.}
Upon new online data points, we update the NRRF with a batch of hybrid data, including online data points and simulated data from the enhanced simulator. 
We denote the hybrid data batch $\mathcal{D}_h$ as 
$\mathcal{D}_h = \mathcal{D}_{on} \cup \mathcal{D}_{sim}$,
where $\mathcal{D}_{on}$ and $\mathcal{D}_{sim}$ are the batch of online and simulated data points.
On the one hand, we build $\mathcal{D}_{sim}$ by sampling proximal locations (e.g., within tens of meters) of the latest online data point in the enhanced simulator. 
This is because the enhanced simulator is updated with the latest online data point by the material tuning algorithm (see Sec.~\ref{sec:augmentation}), i.e., the material of proximal buildings.
The simulator RSRPs at proximal locations could provide the latest information regarding the real-world network. 
On the other hand, we obtain $\mathcal{D}_{on}$ by sampling historical online data points from an experience replay buffer.
We create this replay buffer to store all observed online data points from the real-world network.
We develop the replay buffer to be fixed-size, and remove outdated data points if it is full.
Thus, with the hybrid data batch, the NRRF will be updated online with both the latest online data points and simulated data with embedded domain knowledge.


\section{System Implementation}

\begin{table}[!t]
\centering
\caption{Simulation parameters in Sionna RT.}
\label{tb:TX_RX_config}
\begin{tabular}{c|cc}
\hline\hline
\renewcommand{\arraystretch}{1.6}
\textbf{Parameters}          & \textbf{TX}        &\textbf{RX}        \\ \hline
Number of rows               & 1                  & 1                 \\
Number of columns            & 4                  & 1                 \\
Vertical antenna spacing     & 0.5                & -                 \\
Horizontal antenna spacing   & 0.5                & -                 \\ 
Antenna pattern              & tr38901            & iso             \\
Position                     & (-511.8,-367.1)    & see Fig.~\ref{fig:data_collection_map} \\
Orientation                  & ($\pi/6$,0,0)      & (0,0,0)           \\
Height                       & 34 m               & 1 m               \\\hline\hline
\end{tabular}
\end{table}

\begin{figure}[!t]
	\centering
	\includegraphics[width=3.3in]{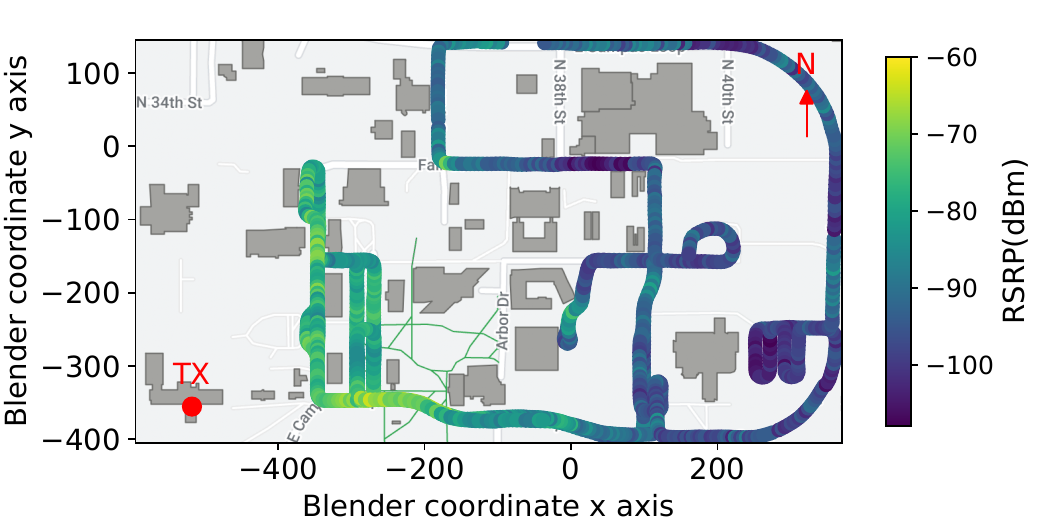}
	\caption{The campus map and trajectory of data collocation.}
	\label{fig:data_collection_map}
\end{figure}

\textbf{The Network.}
We collect online data from a public cellular network, which has one base station located in the lower left corner of Fig.\ref{fig:data_collection_map}. 
The network is in band 66 (LTE FDD) with the downlink radio frequency at $[2110, 2200]$ MHz. 
We use a Quectel RM520N-GL modem with omni-antenna as the UE, which is connected to a Windows 11 laptop via a dongle. 
To collect online data (i.e., RSRP), we mount the UE on the top of a test vehicle and drive the vehicle on campus roads, as shown in Fig.~\ref{fig:data_collection_map}. 
Besides, we use a Real-time Kinematic (RTK) positioning system to collect high-precision GPS locations (centimeter-level precision).
We installed the RTK base station on a rooftop and mounted the RTK receiver next to the UE (fixed distance).
In total, we collected 1.5K online data points and partitioned them into training and testing sets.

\textbf{The Simulator.}
We build a network simulator with Sionna RT (v0.18.0), using Blender 3 and OpenStreetMap3D to create the scene of the campus map. 
First, we create a scene of the target campus area in Blender 3 (Fig.~\ref{fig:data_collection_map}) by importing from OpenStreetMap3D. 
Second, we initialize the default radio material for all buildings, i.e., \textit{itu-marble} for building surfaces and \textit{itu-concrete} for the ground. 
Third, we import the scene in Blender to Sionna RT, and configure all the needed simulation parameters. 
Table~\ref{tb:TX_RX_config} shows the simulation parameters to configure the antenna array of TX and RX, where we tried our best to calibrate these parameters with the real-world network.
When computing the path in Sionna scene, the maximum depth allowed for tracing paths is set to 2, considering both line-of-sight (LoS) and non-line-of-sight (NLOS) paths.
We enable reflection and disable diffraction and scattering during the ray tracing, to trade off accuracy and computation complexity.
Additionally, we adopt the Fibonacci method for the ray tracing computation, with the GPU acceleration (TensorFlow 2.13). 
Besides, Table \ref{tb:link_budget_table} shows additional simulation parameters, such as transmitter power and losses. 
We compute the path loss from the Channel Impulse Responses (CIRs) \cite{hoydis2023learning} calculated by the ray tracer using \textit{sionna.rt.Paths.cir()}, obtaining the discrete complex baseband-equivalent CIR and summing the squared amplitudes of all channel taps.
We use a set of materials recommended in ITU-R P.2040-2 \cite{series2015effects} for material tuning.

\begin{table}[]
\centering
\caption{Additional simulation parameters in Sionna RT.}
\label{tb:link_budget_table}
\begin{tabular}{c|c}
\hline\hline
\renewcommand{\arraystretch}{1.6}
\textbf{Parameters}       & \textbf{Value}                           \\ \hline
 Max. transmit power        & 37 dBm                                   \\
Transmitter antenna gain & 12 dBi                                    \\
Transmitter losses       & 15 dB                                    \\ 
Path loss     & Get from Sionna RT (dB)                     \\
Miscellaneous losses     & 18 dB                                    \\ 
Receiver antenna gain    & 1 dBi                                    \\
Receiver losses          & 18 dB                                  \\\hline\hline
\end{tabular}
\end{table}


\textbf{The Algorithms.}
We implement the oneTwin system on a workstation with Intel i7-14700K (64G RAM), NVIDIA RTX 4090 (24G RAM), and Ubuntu 22.04.
Overall, we implement two algorithms under Python 3.8, PyTorch 2.3, and CUDA 12.1.
On the one hand, we develop the material tuning algorithm with the GP and EI as the surrogate model and acquisition function in the Bayesian optimization, respectively.
We implement GP by using \textit{scikit-learn} toolbox with the \textit{GaussianProcessRegressor} module and \textit{Matern} kernel.
The maximum number of iterations is 25 with 10 warm start (i.e., random sampling). 
On the other hand, we use the Adam optimizer with a learning rate of 1e-3 and a decay rate of 5e-5 in the neural learning algorithm.
The number of iterations in the initial offline training is 2000, with a batch size of 256.
We use $\lambda=0.4$ in the EWC to balance current and previous changes during the online neural learning.
In the NRRF, the attenuation DNN consists of 8 MLP layers, each containing 256 neurons, and the radiance DNN comprises 2 MLP layers, with the first layer having 256 neurons and the second layer having 128 neurons.
During online neural learning, the batch size of hybrid data is 32, which includes 16 online data points and 16 simulated data points.


\section{Performance Evaluation}
\label{sec:evaluation}

 \begin{figure*}[!t] 
\captionsetup{justification=centering}
  \begin{minipage}[t]{0.49\textwidth}
    \centering
    \includegraphics[width=2.8in, height=1.45in]{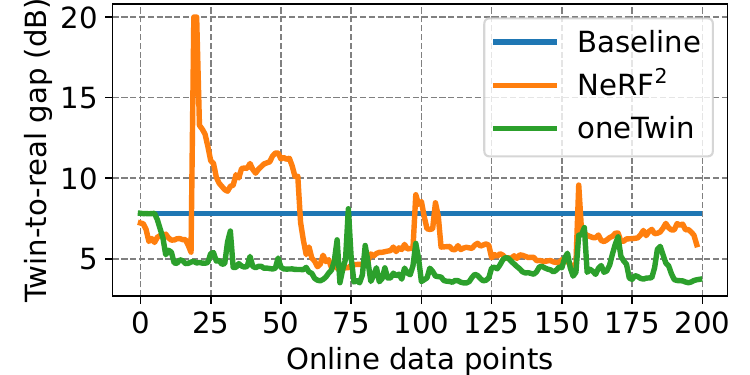}
    \caption{Twin-to-real gap of all systems throughout online digital twinning.}
    \label{fig:core_result}
  \end{minipage}
  \begin{minipage}[t]{0.49\textwidth}
    \centering
    \includegraphics[width=2.8in, height=1.45in]{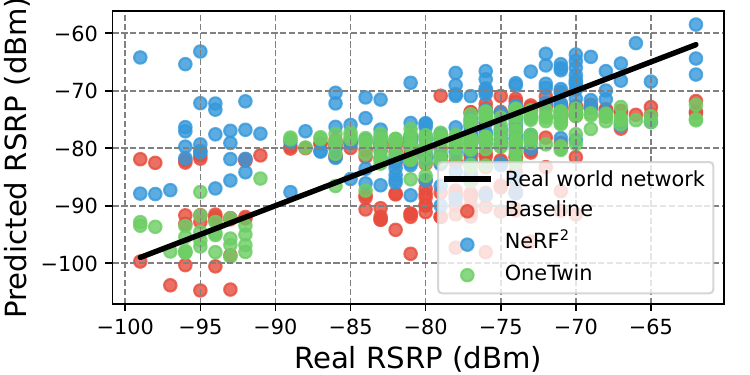}
    \caption{Sample-wise RSRP comparison, twin-to-real gap of oneTwin and NeRF$^2$ are 3.74dB and 5.88dB, respectively.}
    \label{fig:CDF_sys_sim_augsim}
  \end{minipage}
\end{figure*}

We compare oneTwin with the following systems:
\begin{itemize}[leftmargin=*]
    \item \textbf{Baseline}: In Baseline, we adopt Sionna RT \cite{hoydis2022sionna}, a state-of-the-art open-source link-level network simulator, to build the digital network twin via simulator-based approaches. We tried our best to calibrate various simulation parameters (see Table~\ref{tb:TX_RX_config} and Table~\ref{tb:link_budget_table}) with these of the real-world network. Given an online query, Baseline generates the predicted RSRP via executing the network simulation in Sionna RT. 
    \item \textbf{NeRF$^{2}$}: Neural Radio-Frequency Radiance Fields (NeRF$^{2}$) \cite{zhao2023nerf2} extends the NeRF \cite{mildenhall2021nerf} technique from optics to RF domain and serves as the representative solution of neural-based approaches toward digital network twins. To fit into the setting of online learning, we slightly modify its implementation to update its DNNs with the batch of latest 32 online data points. Given an online query, NeRF$^{2}$ generates the predicted RSRP via inferring its DNNs.
\end{itemize}



\subsection{Overall Performance}
\label{sec:eval:subsec:overall}
Fig.~\ref{fig:core_result} shows the twin-to-real gap under different systems during online digital twinning, with data points collected from real-world networks. 
Since Baseline is built on a static network simulator, its twin-to-real gap remains constant and fails to track real-world network dynamics continually.
During online digital twinning, the twin-to-real gap of NeRF$^{2}$ is very unstable, exhibiting sharp and quick variations (e.g., starting at the 20th data point). 
This instability can be attributed to the limited data points (e.g., tens of data points), which are insufficient to train the DNN of NeRF$^{2}$ to be convergent and robust. 
However, NeRF$^{2}$ generally achieves lower twin-to-real gaps than Baseline, indicating its effectiveness in the implicit representation of complex RF signal propagation.
In contrast, oneTwin achieves the lowest twin-to-real gap among these comparison systems by the end of the online data points. 
Since NRRF is trained offline with extensive simulated data, oneTwin consistently obtains lower twin-to-real gaps than Baseline since the very beginning. 
Throughout online digital twinning, oneTwin continuously reduces the twin-to-real gap without any large spikes (e.g., higher gaps than that of Baseline). 
Eventually, oneTwin achieves a final twin-to-real gap of 3.74dB, which is a reduction of 57.50\% and 36.39\% compared to Baseline and NeRF$^{2}$, respectively.

\begin{figure*}[!t] 
\captionsetup{justification=centering}
  \begin{minipage}[t]{0.32\textwidth}
    \centering
    \includegraphics[width=2.25in, height=1.45in]{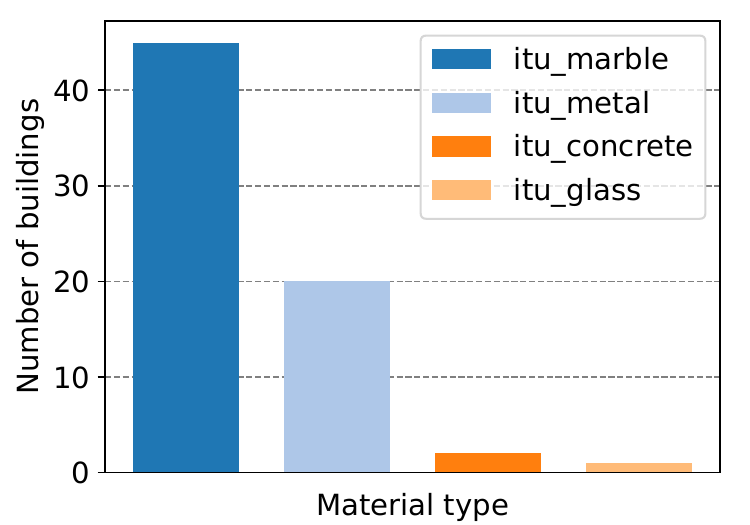}
    \caption{Optimized material of buildings under the oneTwin system.}
    \label{fig:phase1_materials}
  \end{minipage}
  \begin{minipage}[t]{0.32\textwidth}
    \centering
    \includegraphics[width=2.25in, height=1.45in]{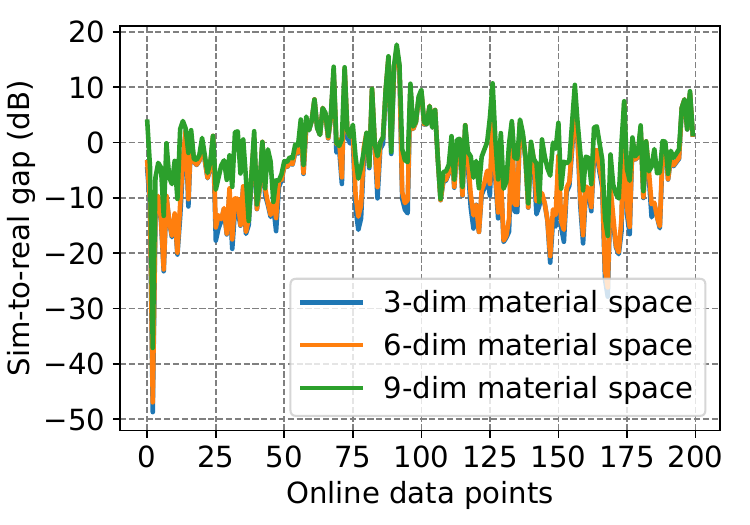}
    \caption{Sim-to-real gap under different material spaces in oneTwin.}
    \label{fig:phase1_diff_material_type_gap_compare}
  \end{minipage}
  \begin{minipage}[t]{0.32\textwidth}
    \centering
    \includegraphics[width=2.25in, height=1.45in]{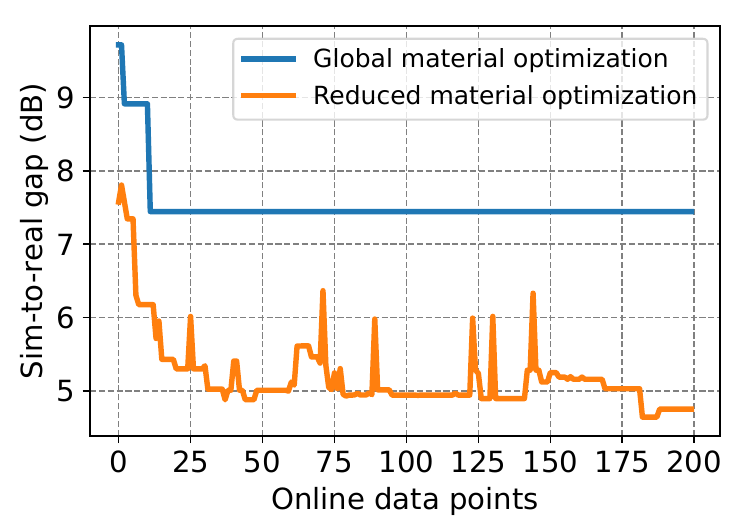}
    \caption{Sim-to-real gap under material optimization strategies in oneTwin.}
    \label{fig:phase1_all_test_gap}
  \end{minipage}
\end{figure*}

\begin{figure*}[!t] 
\captionsetup{justification=centering}
  \begin{minipage}[t]{0.32\textwidth}
    \centering
    \includegraphics[width=2.25in, height=1.45in]{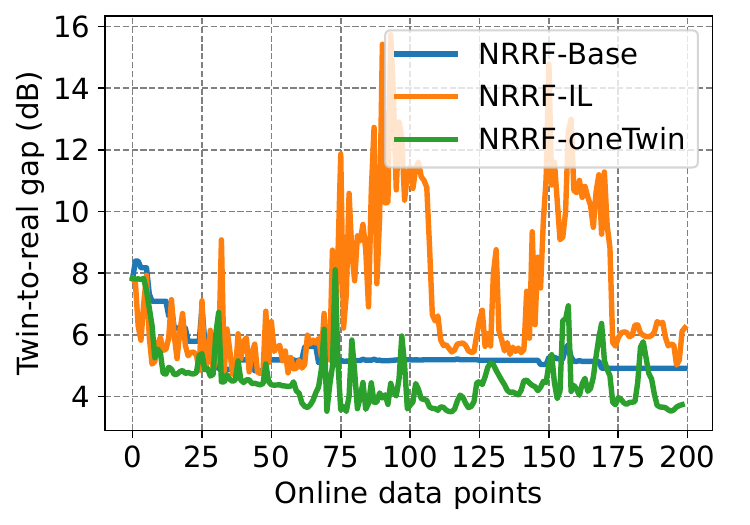}
    \caption{Twin-to-real gap under different NRRF training strategies in oneTwin.}
    \label{fig:phase2_Nerf_result}
  \end{minipage}
  \begin{minipage}[t]{0.32\textwidth}
    \centering
    \includegraphics[width=2.25in, height=1.45in]{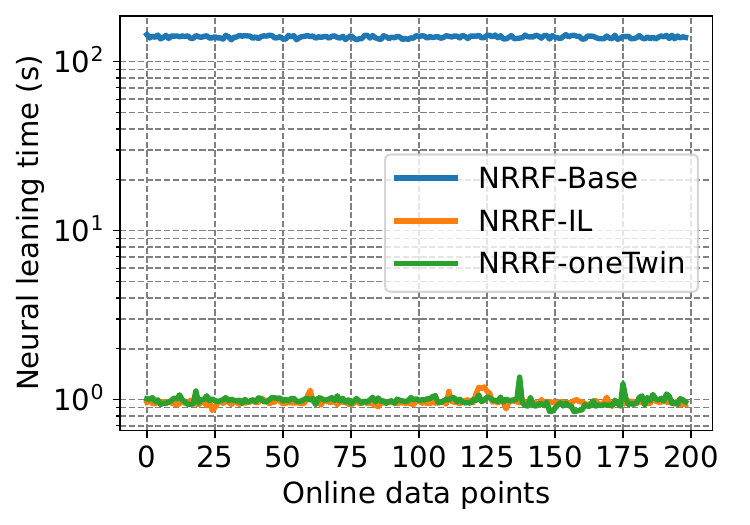}
    \caption{Time consumption under NRRF training strategies in oneTwin.}
    \label{fig:phase2_Nerf_elapsed_time}
  \end{minipage}
  \begin{minipage}[t]{0.32\textwidth}
    \centering
    \includegraphics[width=2.25in, height=1.45in]{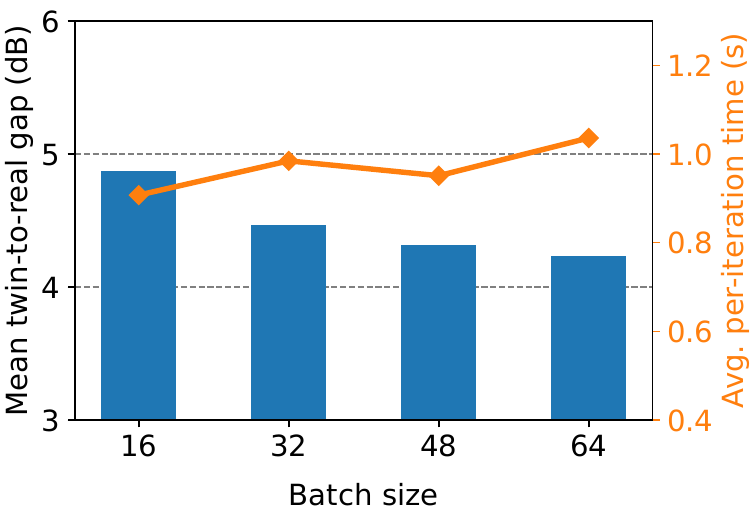}
    \caption{ Mean twin-to-real gap under different batch sizes in oneTwin.}
    \label{fig:phase2_diff_batch_size}
  \end{minipage}
\end{figure*}

In addition, we measure the time consumption of online digital twinning upon the arrival of each online data point. 
NeRF$^{2}$ needs an average of 0.40 seconds, while Baseline takes an average of 0.26 seconds. 
Here, oneTwin needs an average of 0.98 seconds, which is the latency of online neural learning in NRRF, and the latency of material tuning in enhanced simulator (5.8 seconds) is hidden via its asynchronous updating mechanism (see Fig.~\ref{fig:core_sample_efficiency} on performance impacts).
Fig.~\ref{fig:CDF_sys_sim_augsim} shows the sample-wise RSRPs comparison under the test dataset.
Here, the closer to the real-world network curve, the lower the twin-to-real gap, thus better performance.
Given the noticeably distal Baseline curve, we observe that oneTwin is very close to that of the real-world network.
Overall, these results verify the high fidelity and tractability of oneTwin.


\subsection{oneTwin: Enhanced Simulator}
In this subsection, we dissect oneTwin by evaluating the design of enhanced simulator from multiple aspects.

\textbf{Material Space.}
In oneTwin, we define the material space of the material tuning algorithm with 9 kinds of materials.
Fig.~\ref{fig:phase1_materials} shows the statistics of final optimized materials for all 68 buildings, which includes 45 \textit{itu-marble}, 20 \textit{itu-metal}, 1 \textit{itu-glass}, and 1 \textit{itu-concrete}, in addition to the \textit{itu-concrete} ground.
Moreover, we evaluate the impact of the material space on the achievable twin-to-real gap under the oneTwin system.
Fig.~\ref{fig:phase1_diff_material_type_gap_compare} shows the RSRP difference under all 200 data points in the test dataset.
It can be seen that, the larger the material space, the lower the twin-to-real gap (i.e., [8.55, 7.99, 4.75] dB) can be achieved by the oneTwin system.
Because a larger material space could have a higher chance of containing realistic materials (or similar materials) in real-world networks. 
Additionally, we notice that their average time consumption increases gradually (i.e., [2.49, 5.39, 5.80] seconds), which supports incorporating more material types to further reduce the twin-to-real gap, if applicable.




\textbf{Reduced Material Optimization.}
In oneTwin, we design the material tuning algorithm to optimize materials only for the involved objects of the current online data point (see Sec.~\ref{subsec:augmentation-solution}), which aims to reduce the optimization space and accelerate the material tuning.
Fig.~\ref{fig:phase1_all_test_gap} shows the twin-to-real gap under different material optimization strategies in the oneTwin system, throughout online digital twinning.
Given the same configurations, i.e., the same warmup and total iterations, we observe that the time consumption of using global material optimization is 41.03\% more than that of using reduced material optimization.
In addition, the twin-to-real gap (7.44dB) can be substantially worse if using the strategy of global material optimization. 
In contrast, the automatic material tuning algorithm in oneTwin can achieve a 4.75dB twin-to-real gap (i.e., a 36.16\% reduction) by adopting the strategy of reduced material optimization.
Moreover, we observe that the twin-to-real gap may be reduced (from 7.44dB to 6.78dB) if adding more search iterations (i.e., 100 iterations).
However, this would significantly increase its time consumption, which would fail the tractability attribute of oneTwin.

\subsection{oneTwin: Neural Radio Radiance Field}
In this subsection, we dissect the oneTwin system by evaluating the design of neural radio radiance field (NRRF).
We compare the online neural learning algorithm with other comparison baselines, to illustrate the impact of incremental learning and hybrid data points (i.e., simulated and online).
Here, \emph{NRRF-IL} updates the NRRF with incremental learning technique via elastic weight consolidation (EWC), and \emph{NRRF-Base} updates the NRRF with all data points.
Both NRRF-IL and NRRF-Base use data points from sampling the ever-updating enhanced simulator (see Sec.~\ref{sec:augmentation}), without including online data points.

Fig.~\ref{fig:phase2_Nerf_result} shows the twin-to-real gap under the above comparison baselines and the oneTwin system.
We can see that, the NRRF-Base achieves stable convergence performance without significant spikes of twin-to-real gap.
However, because it trains with all sampled data points, its time consumption increases quickly and cannot scale to support an ever-increasing number of data points (e.g., thousands if not more) in real-world networks.
Besides, the twin-to-real gap of NRRF-IL highly varies throughout online digital twinning.
In contrast, the NRRF-oneTwin achieves the lowest twin-to-real gap on average with generally stable convergence performance.
This result suggests that, incorporating online data points can regularize the incremental learning during the update of NRRF in the oneTwin system. 
Moreover, Fig.~\ref{fig:phase2_Nerf_elapsed_time} shows the time consumption of updating NRRF, where NRRF-oneTwin needs 0.98 seconds, NRRF-IL takes 0.97 seconds, and NRRF-Base requires 139.82 seconds.
These results verify the efficiency of the online neural learning algorithm in the oneTwin system.

Fig.~\ref{fig:phase2_diff_batch_size} shows the impact of batch size on the mean twin-to-real gap in the oneTwin system.
Here, we define the metric of \emph{mean twin-to-real gap} as the mean of the twin-to-real gap throughout online digital twinning, which aims to represent the twin-to-real gap statistically.
We can see that, the twin-to-real gap can be generally reduced with higher batch sizes in updating the NRRF in oneTwin.
Besides, their time consumption is very similar, which is attributed to the high computation capacity of GPU (i.e., RTX 4090).
This result suggests that, it would be beneficial to increase the batch size if the capacity of system hardware allows.

\begin{figure*}[!t] 
\captionsetup{justification=centering}
  \begin{minipage}[t]{0.32\textwidth}
    \centering
    \includegraphics[width=2.25in, height=1.45in]{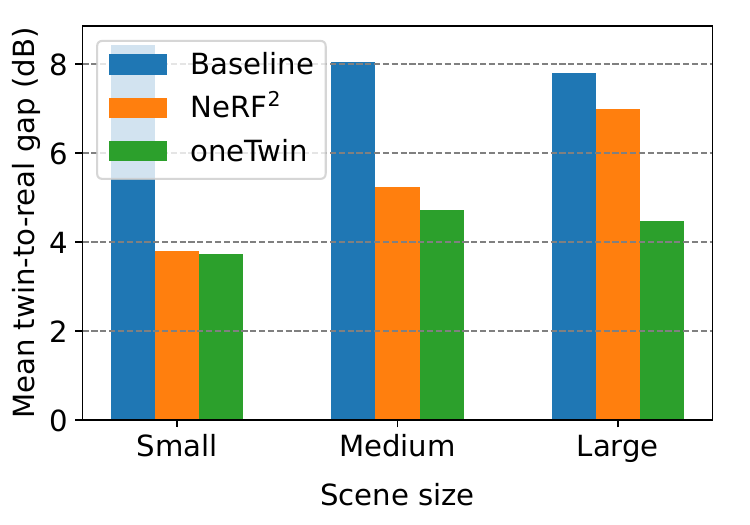}
    \caption{Mean twin-to-real gap under different scene sizes.}
    \label{fig:core_diff_scene_size}
  \end{minipage}
  \begin{minipage}[t]{0.32\textwidth}
    \centering
    \includegraphics[width=2.25in, height=1.45in]{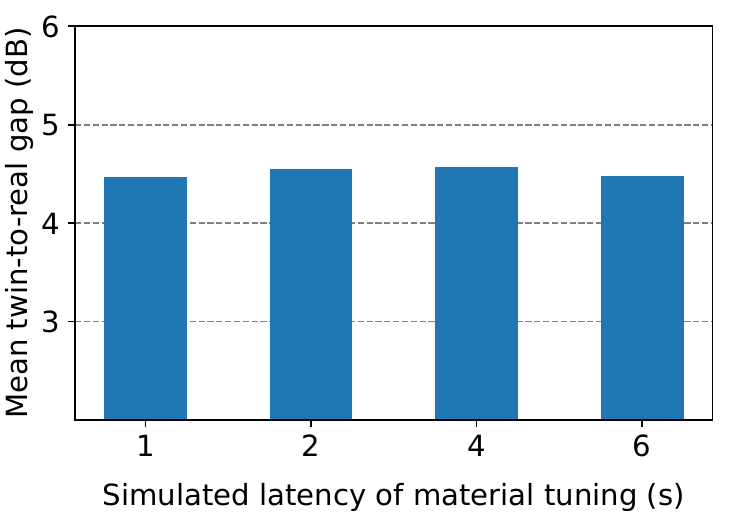}
    \caption{Mean twin-to-real gap under simulated material tuning latencies.}
    \label{fig:core_sample_efficiency}
  \end{minipage}
  \begin{minipage}[t]{0.32\textwidth}
    \centering
    \includegraphics[width=2.25in, height=1.45in]{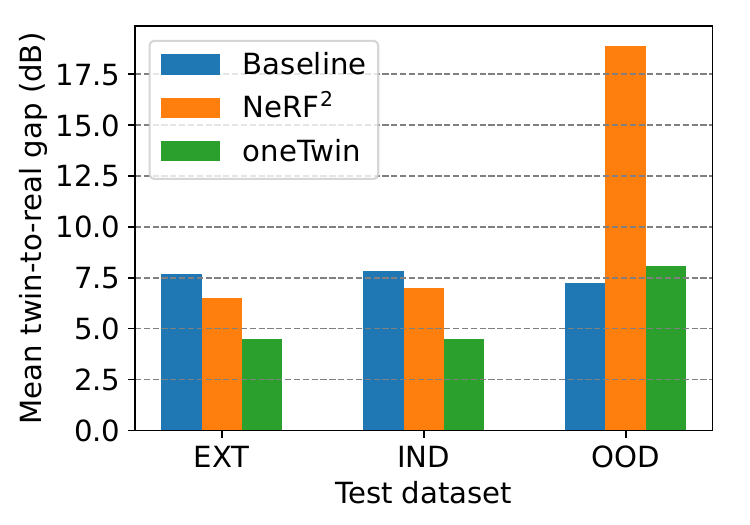}
    \caption{ Mean twin-to-real gap of all systems under different test datasets.}
    \label{fig:core_result_out}
  \end{minipage}
\end{figure*}

\subsection{Scalability, Tractability and Generalizability}
\label{subsec:scale_effi_general}
In this subsection, we evaluate the oneTwin system, in terms of scalability, tractability and generalizability. 

For the evaluation of scalability, we test the oneTwin system under three scenarios with different geographic area sizes.
The largest scene is $1200m\times800m$, with 68 buildings and 400 online data points.
The medium scene is $900m\times550m$, with 39 buildings and 300 online data points. 
The small scene is $650m\times370m$, with 27 buildings and 200 online data points. 
As shown in Fig.~\ref{fig:core_diff_scene_size}, both oneTwin and NeRF$^2$ can achieve a lower mean twin-to-real gap in the smaller scenario, which could be attributed to the reduced size (i.e., less complex) of radio environments.
In addition, the time consumption of updating the NRRF in the oneTwin system generally keeps stable (i.e., [0.78, 0.95, 0.98] seconds). 
This result ensures the high scalability of oneTwin to support large-scale scenarios.



For the evaluation of tractability, we simulate varying latencies introduced by the simulator enhancement, as illustrated in Fig.~\ref{fig:core_sample_efficiency}.
In the current implementation, oneTwin needs 5.80 seconds on average to conduct the material tuning algorithm in the enhanced simulator\footnote{Note that, there are several promising techniques to accelerate the material tuning, such as parallel Bayesian optimization~\cite{wang2016parallel,frazier2018tutorial,de2019sampling} and accelerated execution of Sionna RT.}, which is hidden by the asynchronous updating mechanism.
Specifically, we observe that the execution of Sionna RT requires 0.204 seconds on average, which totals $0.204 \times 25 = 5.1$ seconds during material tuning, and material tuning algorithm itself only takes 0.7 seconds. 
Indeed, this asynchronous updating will lead to delayed tracking of real-time dynamics (i.e., 6 seconds).
However, we observe that, the impact of the asynchronous updating mechanism on the twin-to-real gap difference is minimal.
For example, performing material tuning synchronously results in the mean twin-to-real gap of 4.46 dB, which is only a 0.45\% reduction over the current asynchronous approach (4.48dB). 
This confirms the efficacy of the asynchronous updating mechanism in maintaining the fidelity of the simulator enhancement.


For the evaluation of generalizability, we show the mean twin-to-real gap of all systems in Fig.~\ref{fig:core_result_out}, under different test datasets.
In addition to previous test dataset (denoted as in-distribution (IND)), we build another test dataset with all online training data points (denoted as EXT). 
Moreover, we collect an out-of-distribution (OOD) test dataset from distal geographic areas on campus, whose data points are unseen by all systems.
It can be seen that NeRF$^2$ performs much worse under OOD test dataset than under other datasets, which raises concerns regarding its generalizability. 
In contrast, oneTwin achieves a mean twin-to-real gap of 8.09dB (a 57.20\% reduction compared to NeRF$^2$) under the OOD test dataset.
It is worth mentioning that we observe oneTwin eventually reaches a 5.99dB twin-to-real gap (i.e., lower than that of Baseline), which could be further reduced with more online data points. 
This result verifies that oneTwin can generalize to unseen geographic areas, assuring its high fidelity.





\section{Related Work}

\textbf{Digital Network Twin.}
Digital network twin attracts considerable attention in reshaping and defining the next-generation mobile network~\cite{villa2024colosseum, polese2024colosseum, testolina2024boston, borges2024caviar, zhao2023nerf2}.
Polese \emph{et al.} proposed Colosseum~\cite{polese2024colosseum} as the digital twin for open radio access networks (Open RAN), which leverages the world’s largest wireless network emulator with hardware-in-the-loop to reproduce configuration and topologies in real-world network deployment.
Testolina \emph{et al.} proposed BostonTwin~\cite{testolina2024boston} that combines a high-fidelity 3D model of the city of Boston, MA, and existing geo-spatial data of cellular base station deployment.
However, these works are built based on only network simulators, which suffer from non-trivial simulation-to-reality discrepancy~\cite{liu2022atlas,shi2021adapting} and can hardly be inferred in real-time for large-scale scenarios.
By translating NeRF~\cite{mildenhall2021nerf} into the RF signal domain, Zhao \emph{et al.} proposed NeRF$^2$~\cite{zhao2023nerf2} as the representative deep-learning framework for radio propagation and channel understanding.
NeRF$^2$ shows a great potential to achieve digital network twin, especially in the physical layer, with high fidelity and tractability.
Jia \emph{et al.}~\cite{jia2024neural} studied the combination of traditional ray tracing with the neural reflectance field to learn the complex material reflectivity functions in real-world environments.
However, these works fail to timely adapt to time-evolving real-world dynamics, due to their high training complexity.

\textbf{Online Learning.}
Online learning refers to the process of continually training DNNs, under the stream-like online data, to adapt to time-varying changes (e.g., a new category in classification tasks)~\cite{de2021continual, hadsell2020embracing, liu2021constraint, liu2019direct}.
One of the key considerations is to avoid the phenomenon of catastrophic forgetting~\cite{kirkpatrick2017overcoming}, where the learning of new tasks may overwrite previously learned knowledge.
To address this challenge, multiple techniques have been explored, such as elastic weight consolidation (EWC), experience replay, and meta-learning~\cite{wang2024comprehensive}.
In addition, Wang \emph{et al.}~\cite{wang2023lifelong} proposed a lifelong learning method based on EWC to continually update the digital twin. 
In the domain of network management, several works focused on online learning control policies to adapt to non-stationary networks.
To adapt to time-varying dynamics in network slicing, Liu \emph{et al.}~\cite{liu2022atlas} proposed Atlas, an online network slicing system, that automates the service configuration of slices with assured service level agreements (SLAs).
However, these works learn online with only very limited real-world data points, where their DNN models suffer from the issue of poor generalizability and robustness over continually varying network dynamics.

\section{Conclusion}
In this paper, we presented oneTwin, the first online digital network twin system, to achieve real-time and accurate prediction of physical layer metrics (i.e., RSRP).
We achieved the enhanced simulator with a material tuning algorithm that incrementally updates the building materials in the simulator.
We achieved the NRRF with a neural learning algorithm that updates the NRRF online, based on both online and simulated data.
Results show oneTwin substantially outperforms state-of-the-art solutions in fidelity, synchronicity and generalizability.

\section*{Acknowledgment}
This work is partially supported by the US National Science Foundation under Grant No. 2321699 and No. 2333164. 
We appreciate the support from NVIDIA Academic Grant.

\clearpage
\bibliographystyle{IEEEtran}
\bibliography{ref/reference, ref/qiang}

\end{document}